\documentstyle[12pt,epsf,psfig]{article}

\newcommand{\gtrsim}{ \mathop{}_{\textstyle \sim}^{\textstyle >} }

\begin{document}
\baselineskip 0.7cm

\renewcommand{\thefootnote}{\fnsymbol{footnote}}
\setcounter{footnote}{1}

\begin{titlepage}
\begin{center}

\hfill    IASSNS-HEP-98/94\\
\hfill    hep-ph/9811257\\
\hfill    November, 1998\\

\vskip .5in

{\large \bf Electric Dipole Moments in Gauge Mediated Models 
          \\and a Solution to the SUSY CP Problem}

\vskip .5in

{\large Takeo~Moroi}

\vskip .2in

{\it School of Natural Sciences, Institute for Advanced Study\\
Olden Lane, Princeton, NJ 08540, U.S.A.}

\end{center}

\vskip .5in

\begin{abstract}

The SUSY CP problem in the framework of gauge mediated SUSY breaking
model is considered. We first discuss the electric dipole moments of
the electron and neutron, which are likely to be larger than the
experimental upper bound if all the phases in the Lagrangian are
$O(1)$. We derive a constraint on the phases in the so-called $\mu$-
and $B_\mu$-parameters and gaugino masses. Then, we discuss a model in
which the CP violating phase can be adequately suppressed. If the
$\mu$- and $B_\mu$-parameters originate from the same superpotential
interaction as the SUSY breaking field, the CP violating phase
vanishes.  However, in this class of models, the ratio $B_\mu/\mu$
becomes too large, and we discuss a possible scenario to fix this
problem.

\end{abstract}

\end{titlepage}

\renewcommand{\thefootnote}{\#\arabic{footnote}}
\setcounter{footnote}{0}

\section{Introduction}

Supersymmetry (SUSY) is an attractive solution to one of the most
serious fine-tunings in nature, i.e., it ensures the stability of the
electroweak scale against radiative corrections. However, the SUSY
standard model (SSM) may introduce other (less severe) fine-tunings,
since some of the parameters in the SSM and/or their phases must be
very small to avoid unwanted FCNC and CP violating processes. (These
are called SUSY FCNC problem and SUSY CP problem.)

In gauge mediated SUSY breaking model~\cite{LEGM}, the SUSY FCNC
problem can be beautifully solved. In this scheme, the mechanism to
mediate SUSY breaking to the SSM sector does not distinguish between
flavors, and the universality of the scalar mass matrices is
automatically guaranteed.

However, the SUSY CP problem still remains. In particular, in gauge
mediated model, the electric dipole moments (EDMs) of the electron and
neutron are likely to be larger than the current experimental
constraint if the possible phases in the Lagrangian are all
$O(1)$. Therefore, it is better to come up with some idea to suppress
the CP violating phase in the gauge mediated model.

In the first half of this letter, we discuss the electron and neutron
EDMs in the framework of the gauge mediated model, and derive a
constraint on the CP violating phase. As a result, we will see that
the EDMs are likely to be larger than the current experimental
constraint if the CP violating phase is $O(1)$. Then, in the second
half, we consider a mechanism to suppress this CP violating phase so
that the EDMs are within the experimental constraints.

\section{SUSY CP Problem in Gauge Mediated Model}

First, we discuss constraints on the phases in the gauge mediated
model. In the gauge mediated model, all the off-diagonal elements in
the sfermion mass matrices vanish. Furthermore, so-called
$A$-parameters are not generated at the one loop level. Therefore, CP
violating phases in these parameters are suppressed enough to be
consistent with experimental constraints.

However, (some combinations of) the phases in the gaugino masses,
$\mu$-parameter, and $B_\mu$-parameter are physical, and in general,
they can be large enough to conflict with experimental constraints. In
particular, since the mechanism to generate $\mu$- and
$B_\mu$-parameters are unknown, there is no guarantee of cancellation
between their phases.

Let us discuss this issue in more detail. The relevant part of the
Lagrangian of the SSM can be written as
 \begin{eqnarray}
  {\cal L} = - \int d^2\theta \mu H_1 H_2
  - B_\mu H_1 H_2 
  - \frac{1}{2} \left( m_{G1} \tilde{B} \tilde{B}
  + m_{G2} \tilde{W} \tilde{W} + m_{G3} \tilde{G} \tilde{G} \right)
  + {\rm h.c.}
 \end{eqnarray}
 Here, $H_1$ and $H_2$ are the Higgs fields coupled to the down-type
and up-type quarks, and $\tilde{B}$, $\tilde{W}$, and $\tilde{G}$ are
the gauginos for U(1)$_{\rm Y}$, SU(2)$_{\rm L}$, and SU(3)$_{\rm C}$
gauge groups, respectively. In the above Lagrangian, all the
parameters $\mu$, $B_\mu$, and $m_{Gi}$ can be complex. However, by
using phase rotations of Higgs bosons, Higgsinos, and gauginos, we can
make some of them real. To be more specific, denoting\footnote
 {In gauge mediated model, phases of the gaugino masses are
universal, and we denote them as $\theta_G$.}
 \begin{eqnarray}
  \mu = e^{i\theta_\mu} |\mu|, ~~~
  B_\mu = e^{i\theta_B} |B_\mu|, ~~~
  m_{Gi} = e^{i\theta_G} |m_{Gi}|,
 \end{eqnarray}
 physical quantities depend only on the combination
 \begin{eqnarray}
  \theta_{\rm phys} \equiv
  {\rm Arg}(\mu B^*_\mu m_{G})
  = \theta_\mu - \theta_B + \theta_G.
 \end{eqnarray}

In the gauge mediated model, flavor symmetries are well preserved in
the squark and slepton mass matrices, and SUSY contributions to the CP
violation in FCNC processes are negligible. However, as discussed in
many works~\cite{edm_susy}, the EDMs of the electron and neutron are
important check points. Indeed, non-vanishing $\theta_{\rm phys}$ may
induce sizable EDMs.

 \begin{figure}[t]
 \centerline{\epsfxsize=0.55\textwidth\epsfbox{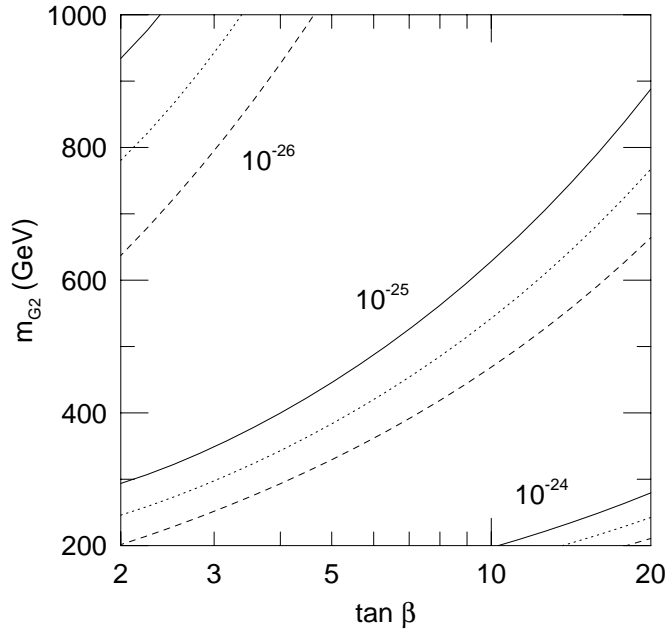}}
 \caption{Contours of the constant electron EDM in gauge mediated
model on $\tan\beta$ vs.~$m_{G2}$ plane. Contours are
$|d_e|/e=10^{-26}$, $10^{-25}$, and $10^{-24}$~cm, from above. Here,
we take $\sin\theta_{\rm phys}=1$, $N_5=1$, and $M_{\rm
mess}=10^{5}$~GeV (solid lines), $10^{10}$~GeV (dotted lines), and
$10^{15}$~GeV (dashed lines).}
 \label{fig:de}
 \end{figure}

In order to discuss the constraint on $\theta_{\rm phys}$, we
calculate the electron EDM $d_e$ in the framework of the gauge
mediated model for several values of the messenger scale $M_{\rm
mess}$. In the calculation, we take $\sin\theta_{\rm phys}=1$ and
$N_5=1$, where $N_5$ is the number of the vector-like messenger
multiplet in units of ${\bf \bar{5}}+{\bf 5}$ representation of
SU(5)$_{\rm G}$. The result is shown in Fig.~\ref{fig:de}. One should
note that $d_e$ is proportional to $\sin\theta_{\rm phys}$, and hence
we can estimate $d_e$ for other values of $\sin\theta_{\rm phys}$ by
rescaling the result given in Fig.~\ref{fig:de}. Furthermore, the EDM
of the electron is enhanced for larger values of $\tan\beta\equiv
\langle H_2\rangle/\langle H_1\rangle$. The mechanism of this
enhancement is the same as those for other leptonic penguin diagrams
such as for the muon magnetic dipole moment~\cite{g-2,PRL79-4752} and
for lepton-flavor violating processes~\cite{LFV}. In particular, the
electron EDM comes from diagrams which are very similar to those for
the muon $g-2$, and those quantities are closely related in the gauge
mediated model:
 \begin{eqnarray}
  d_e \simeq \frac{m_e}{2m_\mu^2} \tan\theta_{\rm phys} 
  \times a^{\rm SSM}_\mu,
 \end{eqnarray}
 where $a^{\rm SSM}_\mu=\frac{1}{2}(g_\mu-2)^{\rm SSM}$ is the SSM
contribution to the muon magnetic dipole moment.

The experimental constraint on the electron EDM is remarkably good. By
using $d_e=(0.18\pm0.12\pm0.10)\times 10^{-26}e$~cm~\cite{PRA50-2960},
we obtain a constraint on the electron EDM:
 \begin{eqnarray}
  |d_e| \leq 0.44 \times 10^{-26} e ~{\rm cm},
 \label{de_exp}
 \end{eqnarray}
 where the right-hand side is the upper bound on $d_e$ at 90~\%~C.L.

With the above constraint, we can derive a bound on $\theta_{\rm
phys}$. Since $d_e$ is proportional to $\sin\theta_{\rm phys}$, the
upper bound on $|\sin\theta_{\rm phys}|$ is given by $10^{-1}$ to
$10^{-3}$, depending on the mass scale of the superparticles and
$\tan\beta$.  If we adopt relatively large value of the wino mass
($m_{G2}\gtrsim 400~{\rm GeV} - 1~{\rm TeV}$, depending on
$\tan\beta$), $\theta_{\rm phys}$ can be as large as 0.1, and it may
not be a serious fine tuning. However, in this case, squarks and
gluino become relatively heavy, and we may lose the motivation for low
energy SUSY as a solution to the naturalness problem. On the other
hand, if we consider lighter wino, $\theta_{\rm phys}$ is constrained
to be less than $O(10^{-2})$, which requires more fine tuning of this
phase. In the following, we consider a solution to this
problem.\footnote
 {In the gauge mediated model, the SUSY CP problem can be solved if
$B_\mu$ vanishes at the messenger scale. For this approach, see
Refs.~\cite{NPB501-297,PRL79-4752}.}

Before discussing the model to suppress $\theta_{\rm phys}$, we
briefly comment on the constraint from the neutron EDM $d_n$.  We can
also obtain a constraint on $\theta_{\rm phys}$ from the neutron
EDM. However, the constraint is less severe because the experimental
constraint on $d_n$ is not as stringent as that on $d_e$, and also
because the heavier squark masses suppress the theoretical value of
$d_n$. With the same underlying parameters, the constraint on
$\theta_{\rm phys}$ is about a few times weaker from $d_n$ than from
$d_e$.

\section{Toy Model and Basic Idea}

Let us consider a toy model in which $\theta_{\rm phys}$ vanishes
automatically.

We denote $X$ as the SUSY breaking field whose scalar and
$F$-components acquire non-vanishing vacuum expectation values
(VEVs). Furthermore, $q$ and $\bar{q}$ are the vector-like messenger
fields, and SUSY breaking parameters in the SSM sector are generated
by integrating them out. In this section, we do not specify the
mechanism that generates an $F$-component for $X$, and we just adopt
the following form of the superpotential:
 \begin{eqnarray}
  W = X F_X^* + y_q X \bar{q} q
  + \frac{y_H}{M_*^{n-1}} X^n H_1 H_2,
 \end{eqnarray}
 where $F_X$ is the VEV of the $F$-component for $X$, $y_q$ and $y_H$
are complex parameters, $n$ is a fixed integer, and $M_*\simeq
2.4\times 10^{18}~{\rm GeV}$ is the reduced Planck scale.

With the above superpotential, $\mu$, $B_\mu$, and $m_{Gi}$ are given
by
 \begin{eqnarray}
  \mu &=& \frac{y_H}{M_*^{n-1}} \langle X^n \rangle,
 \\
  B_\mu &=& 
  \frac{n y_H}{M_*^{n-1}} \langle X^{n-1} \rangle F_X,
 \\
  m_{Gi} &=& \frac{g_i^2}{16\pi^2} c_i N_5 
  \frac{F_X}{\langle X \rangle},
 \label{mG_toy}
 \end{eqnarray}
 where $g_i$ is the relevant gauge coupling constant for the standard
model gauge group and $c_i$ is the group theoretical factor. From
these expressions, we can easily see $\theta_{\rm phys}$ vanishes. In
other words, all the phases in the Lagrangian can be eliminated with
phase rotations of the scalars, chiral fermions, and
gauginos. Therefore, in this case, there is no CP violation in the SSM
(except for the phase in the KM matrix).

However, this scenario is not phenomenologically viable, since the
relative size of the $\mu$- and $B_\mu$-parameters is not in the
required range. The ratio of $B_\mu$ to $\mu$ is given by
 \begin{eqnarray}
  \frac{B_\mu}{\mu} = \frac{nF_X}{\langle X\rangle}.
 \label{B/mu}
 \end{eqnarray}
 On the other hand, if $N_5\sim O(1)$, mass scale of the
superparticles in the SSM sector is estimated as
 \begin{eqnarray}
  m_{\rm SSM} \sim \frac{g_{\rm SM}^2}{16\pi^2} 
  \left| \frac{F_X}{\langle X\rangle} \right|,
 \end{eqnarray}
 where $g_{\rm SM}$ is the relevant gauge coupling constant of the
standard model gauge groups. Then, the ratio $F_X/\langle X\rangle$
has to be of the order of 10 $-$ 100~TeV, where the lower bound is
from the experimental constraint on the masses of the superparticles
while the upper bound is from the naturalness point of view. As a
result, the ratio given in Eq.~(\ref{B/mu}) is about 2 $-$ 3 orders of
magnitude larger than the phenomenologically acceptable
value~\cite{LEGM}.\footnote
 {If the messenger multiplets have large multiplicity of $N_5\sim
100$, $F_X/\langle X\rangle$ can be smaller (see Eq.~(\ref{mG_toy}))
and $B_\mu/\mu$ may be in the required range. Even if $N_5\sim 100$,
perturbative picture can be valid up to the Planck scale if the
messenger scale is as high as $O(10^{16}~{\rm GeV})$. In this case,
the SUSY breaking scalar masses are significantly suppressed at the
messenger scale, but can be generated by the running effect}
 Therefore, this toy model does not work although it has the
attractive feature of vanishing CP violating phase $\theta_{\rm
phys}$. 

\section{Improved Model}

Now, we propose an improved model in which the ratio $B_\mu/\mu$ can
be in the required range. One possibility to suppress the ratio
$B_\mu/\mu$ is to introduce another field which also acquires a
VEV. If this new field couples to the Higgs fields, and also if it can
generate a large enough $\mu$-parameter, the ratio $B_\mu/\mu$ may be
in the required range. Of course, if the VEV of this new field has an
arbitrary phase, the SUSY CP problem cannot be solved. Therefore, the
new field has to be somehow related to the original SUSY breaking
field $X$.

In our model, we duplicate the SUSY breaking sector, and couple both
of them to the Higgs fields. Then, if the SUSY breaking field in one
sector has a larger VEV than the other, the $\mu$-parameter is
enhanced and the ratio $B_\mu/\mu$ can have a required value.
Furthermore, in order not to introduce a new phase which may spoil the
cancellation in $\theta_{\rm phys}$, we impose symmetry which
interchanges these two sectors. If this symmetry is exact, however,
the hierarchy between the VEVs of the two SUSY breaking fields cannot
be generated. Therefore, we introduce a (small) breaking parameter of
this symmetry.

There are two conflicting requirements on the breaking parameter.
First, this breaking parameter has to be large enough so that the VEV
of one SUSY breaking field is about 2 $-$ 3 orders of magnitude
enhanced relative to the other. On the contrary, if this breaking
parameter is too large, its CP violating phase may spoil the
cancellation in $\theta_{\rm phys}$.

It is non-trivial to generate a large enough hierarchy with such a
small breaking parameter. If the VEV of the SUSY breaking field is
determined by the inverted hierarchy mechanism~\cite{PLB105-267},
however, small modifications of the parameters may significantly
change the VEV of the SUSY breaking field. In particular, in this
class of models~\cite{MurDimDva,NPB510-12}, the potential of the SUSY
breaking field is lifted only logarithmically, and a small
perturbation at the Planck scale can result in a significant change of
the minimum of the potential. In this section, we use a simple model
as an example, and see how the scenario mentioned above can work.

 \begin{table}[t]
 \begin{center}
 \begin{tabular}{cccccccc}
 \hline\hline
  {} & 
  {SU(2)$_{\rm B1}$} & {SU(2)$_{\rm B2}$} & {SU(2)$_{\rm S}$} &
  {${\rm SU(2)'_{B1}}$} & {${\rm SU(2)'_{B2}}$} & 
  {${\rm SU(2)'_{S}}$} &
  {SU(5)$_{\rm G}$}
 \\ \hline
  {$\Sigma$} & 
  {${\bf 2}$} & {${\bf 2}$} & {${\bf 1}$} & 
  {${\bf 1}$} & {${\bf 1}$} & {${\bf 1}$} & 
  {${\bf 1}$}
 \\
  {$Q$} & 
  {${\bf 2}$} & {${\bf 1}$} & {${\bf 2}$} & 
  {${\bf 1}$} & {${\bf 1}$} & {${\bf 1}$} & 
  {${\bf 1}$}
 \\
  {$\bar{Q}$} & 
  {${\bf 1}$} & {${\bf 2}$} & {${\bf 2}$} & 
  {${\bf 1}$} & {${\bf 1}$} & {${\bf 1}$} & 
  {${\bf 1}$}
 \\
  {$q_5$} & 
  {${\bf 2}$} & {${\bf 1}$} & {${\bf 1}$} & 
  {${\bf 1}$} & {${\bf 1}$} & {${\bf 1}$} & 
  {${\bf 5}$}
 \\
  {$\bar{q}_5$} & 
  {${\bf 1}$} & {${\bf 2}$} & {${\bf 1}$} & 
  {${\bf 1}$} & {${\bf 1}$} & {${\bf 1}$} & 
  {${\bf \bar{5}}$}
 \\
  {$q_1$} & 
  {${\bf 2}$} & {${\bf 1}$} & {${\bf 1}$} & 
  {${\bf 1}$} & {${\bf 1}$} & {${\bf 1}$} & 
  {${\bf 1}$}
 \\
  {$\bar{q}_1$} & 
  {${\bf 1}$} & {${\bf 2}$} & {${\bf 1}$} & 
  {${\bf 1}$} & {${\bf 1}$} & {${\bf 1}$} & 
  {${\bf 1}$}
 \\
  {$\Sigma'$} & 
  {${\bf 1}$} & {${\bf 1}$} & {${\bf 1}$} & 
  {${\bf 2}$} & {${\bf 2}$} & {${\bf 1}$} & 
  {${\bf 1}$}
 \\
  {$Q'$} & 
  {${\bf 1}$} & {${\bf 1}$} & {${\bf 1}$} & 
  {${\bf 2}$} & {${\bf 1}$} & {${\bf 2}$} & 
  {${\bf 1}$}
 \\
  {$\bar{Q}'$} & 
  {${\bf 1}$} & {${\bf 1}$} & {${\bf 1}$} & 
  {${\bf 1}$} & {${\bf 2}$} & {${\bf 2}$} & 
  {${\bf 1}$}
 \\
  {$q'_5$} & 
  {${\bf 1}$} & {${\bf 1}$} & {${\bf 1}$} & 
  {${\bf 2}$} & {${\bf 1}$} & {${\bf 1}$} & 
  {${\bf 5}$}
 \\
  {$\bar{q}'_5$} & 
  {${\bf 1}$} & {${\bf 1}$} & {${\bf 1}$} & 
  {${\bf 1}$} & {${\bf 2}$} & {${\bf 1}$} & 
  {${\bf \bar{5}}$}
 \\
  {$q'_1$} & 
  {${\bf 1}$} & {${\bf 1}$} & {${\bf 1}$} & 
  {${\bf 2}$} & {${\bf 1}$} & {${\bf 1}$} & 
  {${\bf 1}$}
 \\
  {$\bar{q}'_1$} & 
  {${\bf 1}$} & {${\bf 1}$} & {${\bf 1}$} & 
  {${\bf 1}$} & {${\bf 2}$} & {${\bf 1}$} & 
  {${\bf 1}$}
 \\ \hline\hline
 \end{tabular}
 \caption{Particle content of the model.}
 \label{table:example}
 \end{center} 
 \end{table}

In our discussion, we use a model based on ${\rm [SU(2)]^3\times
[SU(2)']^3\times SU(5)_{G}}$ symmetry as an example, where the
standard model gauge group ${\rm SU(3)_C\times SU(2)_L\times U(1)_Y}$
is embedded in ${\rm SU(5)_{G}}$ in the usual manner. (For the
original SUSY breaking model based on the inverted hierarchy mechanism
with ${\rm [SU(2)]^3\times SU(5)_{G}}$, see Ref.~\cite{NPB510-12}.) We
show the particle content of this model in Table~\ref{table:example}.
Here, ${\rm SU(2)_{S}}$ and ${\rm SU(2)'_{S}}$ are strong gauge
interactions which break supersymmetry, while ${\rm SU(2)_{B}}$'s are
introduced to stabilize the potentials for the SUSY breaking fields.

Assuming a symmetry which interchanges the ${\rm [SU(2)]^3}$ and ${\rm
[SU(2)']^3}$ sectors (which we call $Z_2^{X\leftrightarrow X'}$
symmetry), the superpotential has the following form:
 \begin{eqnarray}
  W &=& y_Q \Sigma \bar{Q} Q + y_5\Sigma \bar{q}_5 q_5
  + y_1 \Sigma \bar{q}_1 q_1
 \nonumber \\ &&
  + y_Q (1+\epsilon_Q)\Sigma' \bar{Q}' Q' 
  + y_5 (1+\epsilon_5) \Sigma' \bar{q}'_5 q'_5
  + y_1 (1+\epsilon_1) \Sigma' \bar{q}'_1 q'_1
 \nonumber \\ &&
  + \frac{y_H}{M_*} {\rm det} \Sigma H_1 H_2
  + \frac{y_H}{M_*} (1+\epsilon_H) {\rm det} \Sigma' H_1 H_2,
 \end{eqnarray}
 where the $\epsilon$'s are the breaking parameters of
$Z_2^{X\leftrightarrow X'}$. If all $\epsilon$'s vanish, there is a
$Z_2^{X\leftrightarrow X'}$ symmetry. 

The symmetry breaking parameters $\epsilon$'s may arise from a VEV of
a field $\phi$ which transforms as $\phi\rightarrow -\phi$ under
$Z_2^{X\leftrightarrow X'}$, for example. If $\phi$ has a coupling
like
 \begin{eqnarray}
  W \sim
  y_1 (\Sigma\bar{q}_1 q_1 + \Sigma' \bar{q}'_1 q'_1)
  + \frac{\phi}{M_*}
  (\Sigma\bar{q}_1 q_1 - \Sigma' \bar{q}'_1 q'_1),
 \label{W_phi}
 \end{eqnarray}
 a small $\epsilon_1$ can be generated if $\phi$ acquires a VEV
smaller than $y_1M_*$. Similar arguments hold for other breaking
parameters. Here, we do not specify the origin of the symmetry
breaking, and just assume they are somehow generated at the Planck
scale.\footnote
 {For example, a quantum modified constraint can induce a VEV of the
symmetry breaking field $\phi$~\cite{PRD56-7183}.}

In this model, SUSY is dynamically broken because of the quantum
modified constraint~\cite{IzaYanInt}. Concentrating on the flat
direction parametrized as $\Sigma\sim{\rm diag}(X,X)$ and
$\Sigma'\sim{\rm diag}(X',X')$, the superpotential
becomes~\cite{NPB510-12}
 \begin{eqnarray}
  W &=& y_Q \Lambda^2 X + y_5 X \bar{q}_5 q_5
  + y_1 X \bar{q}_1 q_1
 \nonumber \\ &&
  + y_Q (1+\epsilon_Q) \Lambda^{\prime 2} X'
  + y_5 (1+\epsilon_5) X' \bar{q}'_5 q'_5
  + y_1 (1+\epsilon_1) X' \bar{q}'_1 q'_1
 \nonumber \\ &&
  + \frac{y_H}{M_*} X^2 H_1 H_2
  + \frac{y_H}{M_*} (1+\epsilon_H) X^{\prime 2} H_1 H_2,
 \label{L_Xqq}
 \end{eqnarray}
 where $\Lambda$ and $\Lambda'$ are the strong scales of ${\rm
SU(2)_{S}}$ and ${\rm SU(2)'_{S}}$, respectively.  Due to
$Z_2^{X\leftrightarrow X'}$, we adopt $\Lambda =\Lambda'$. Because of
the $\Lambda^2 X$ and $\Lambda^{\prime 2} X'$ terms in the
superpotential, $X$ and $X'$ have VEVs in $F$-components and the SUSY
is broken.

Once $X$ and $X'$ acquire VEVs, three important parameters are given
by
 \begin{eqnarray}
  \mu &=& \frac{y_H}{M_*} \langle X^2 \rangle
  + \frac{y_H}{M_*} \langle X^{'2} \rangle (1+\epsilon_H),
 \label{mu_model}
 \\
  B_\mu &=& 
  \frac{2y_H}{M_*} F_X \langle X\rangle
  + \frac{2y_H}{M_*} 
  F_X \langle X'\rangle (1+\epsilon_H) (1+\epsilon_Q^*),
 \label{B_model}
 \\
  m_{Gi} &=& 
  \frac{g_i^2}{8\pi^2} c_i \frac{F_X}{\langle X\rangle}
  + \frac{g_i^2}{8\pi^2} c_i \frac{F_X}{\langle X'\rangle}
  (1+\epsilon_Q^*).
 \label{mG_model}
 \end{eqnarray}
 Then, denoting
 \begin{eqnarray}
  v\equiv |\langle X\rangle |,~~~
  v'\equiv |\langle X'\rangle |,
 \end{eqnarray}
 hierarchy between $v$ and $v'$ can make the ratio $B_\mu/\mu$ to be
in the required range. This is because, for $v\ll v'$, $\mu$- and
$B_\mu$-parameters are dominated by the second term, while the gaugino
mass is determined by the first one. Adopting, for example, $|F_X|\sim
(10^6~{\rm GeV})^2$, $v\sim 10^8~{\rm GeV}$, $y_H\sim 1$, and
$v'/v\sim 10^2 - 10^3$, all the parameters in the SSM sector are in
the required range. In the following, we see how the VEVs and their
large hierarchy are generated.

At the tree level, the potential for the SUSY breaking fields are
completely flat and the minimum of the potential is
undetermined. However, once we consider the wave function
renormalization of the SUSY breaking fields, the potential has a
minimum. Denoting the wave function renormalization for $\Sigma$ and
$\Sigma'$ as $Z_\Sigma$ and $Z_{\Sigma'}$, respectively, the potential
is given by
 \begin{eqnarray}
  V = \frac{|F_X|^2}{Z_\Sigma} + \frac{|F_{X'}|^2}{Z_{\Sigma'}},
 \end{eqnarray}
 where
 \begin{eqnarray}
  F_X^* = y_Q \Lambda^2,~~~
  F_{X'}^* = y_Q (1+\epsilon_Q) \Lambda^2
  = (1+\epsilon_Q) F_X^*.
 \end{eqnarray}
 Therefore, the potential for the SUSY breaking field has a minimum
when $Z_\Sigma$ and $Z_{\Sigma'}$ are maximized. The minimum of the
potential can be estimated by using the renormalization group
equations (RGEs). In our discussion, for simplicity, we take account
of the effect of $g_{\rm B1}$ and $y_1$ with $g_{\rm B1}$ being the
gauge coupling constant for SU(2)$_{\rm B1}$, and neglect the effects
of other coupling constants. This approximation is motivated by the
fact that $y_1$ plays the most important role among the Yukawa
coupling constants in determining the minimum of the potential (see
Ref.~\cite{NPB510-12}). Then, $v=|\langle X\rangle|$ is determined by
solving
 \begin{eqnarray}
  \frac{3}{2} g_{\rm B1}^2(v) - y_1^2(v) = 0.
 \label{2225_vac}
 \end{eqnarray}
 Similar argument holds for the potential of $X'$.

\begin{figure}[t]
 \centerline{\epsfxsize=0.55\textwidth\epsfbox{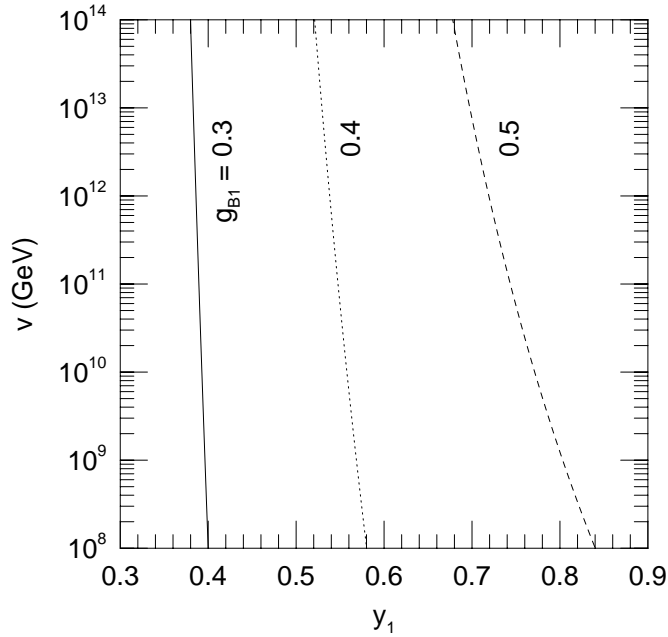}}
 \caption{$v\equiv |\langle X\rangle |$ as a functions of
$y_1(M_*)$. $g_{\rm B1}(M_*)$ is taken to be 0.3 (solid), 0.4
(dotted), and 0.5 (dashed).}
 \label{fig:vevX}
 \end{figure}

Since the scale dependence of the gauge and Yukawa coupling constants
are logarithmic, small modification of the boundary condition at the
Planck scale may result in a significant shift of the minimum of the
potential. In our analysis, we solve the RGEs numerically to see how
the minimum depends on the boundary conditions. For this purpose, we
first fix $g_{\rm B1}$ and $y_1$ at the reduced Planck scale
$M_*$. Then, neglecting other coupling constants, we run them down to
the low energy scale and find the scale $v$ where $Z_\Sigma$ is
maximized (i.e., the VEV of $X$). In Fig.~\ref{fig:vevX}, we show $v$
as a function of $y_1(M_*)$ for several values of $g_{\rm
B1}(M_*)$. As one can see, the VEV of $X$ is sensitive to $y_1(M_*)$,
and small modification of $y_1(M_*)$ results in a large shift of the
minimum of the potential. Fig.~\ref{fig:vevX} shows that $v$ and $v'$
can differ by 2 $-$ 3 orders of magnitude with a small breaking
parameter of $\epsilon_1\sim 10^{-2}-10^{-1}$, depending on $g_{\rm
B1}$. Notice that $\ln (v'/v)\sim\ln[(F_X/\langle
X\rangle)/(B_\mu/\mu)]$ is approximately proportional to
$\epsilon_1$. For example, for $10^2\leq v'/v\leq 10^3$, $\epsilon_1$
is required to be $0.02\leq\epsilon_1\leq 0.03$
($0.04\leq\epsilon_1\leq 0.06$, $0.08\leq\epsilon_1\leq 0.12$) for
$g_{\rm B1}(M_*)=0.3$ (0.4, 0.5). Therefore, in order to make the
ratio $(F_X/\langle X\rangle)/(B_\mu/\mu)$ of the order of
$10^2-10^3$, $\epsilon_1$ has to be mildly tuned at $\sim 50~\%$
level. We believe this is not a serious fine tuning.

Furthermore, most importantly, $\theta_{\rm phys}$ becomes suppressed
in this model. In order to see this suppression, we have to know the
phases of $\langle X\rangle$ and $\langle X'\rangle$. So far, the
phases of $\langle X\rangle$ and $\langle X'\rangle$ are not
determined, since they are related to the $R$-symmetry.  In
supergravity models, however, a constant term exists in the
superpotential to cancel the cosmological constant. This constant term
does not respect the $R$-symmetry and fixes the
phases~\cite{NPB426-3}. The supergravity contributions to the
potential are written as
 \begin{eqnarray}
  V_{\rm SUGRA} = A_Q F_X^* X 
  + A_Q F_{X'}^* (1+\epsilon_A) X' + {\rm h.c.},
 \end{eqnarray}
 where $A_Q$ is a complex SUSY breaking parameter which is of the
order of the gravitino mass. With this potential, for
example, the phase of $\langle X\rangle$ is determined so that the
combination $A_QF_X^*\langle X\rangle$ becomes real. Then, the relative
phase of $\langle X\rangle$ and $\langle X'\rangle$ is given by
 \begin{eqnarray}
  {\rm Arg} \left(\frac{\langle X'\rangle}{\langle X\rangle}\right)
  \simeq
  {\rm Im} (\epsilon_Q^* + \epsilon_A^*).
 \label{Arg(X'/X)}
 \end{eqnarray}
 Therefore, these VEVs are almost aligned irrespective of their
absolute values.

By using Eq.~(\ref{Arg(X'/X)}), $\theta_{\rm phys}$ is calculated as
 \begin{eqnarray}
  \theta_{\rm phys} \simeq {\rm Im}\epsilon_A^*.
 \label{th_phys(eps)}
 \end{eqnarray}
 Therefore, with the current constraint (\ref{de_exp}), the electron
EDM can be suppressed enough for mild values of $\tan\beta$ (less than
about 10) with $\epsilon_A\sim O(10^{-2})$ (see
Fig.~\ref{fig:de}). Even if all the breaking parameters are of the
same order, $\epsilon\sim O(10^{-2})$ can induce large enough
$v'/v$. For larger value of $\tan\beta$, $\epsilon_A$ as small as
$O(10^{-3})$ is required. 

In fact, $\theta_{\rm phys}$ depends only on $\epsilon_A$ as shown in
Eq.~(\ref{th_phys(eps)}), while $\epsilon_1$ plays the most important
role in shifting the VEV. Therefore, if the breaking parameters may
have hierarchy, requirements on the model are more relaxed. In
particular, in the framework of supergravity, $\epsilon_A$ vanishes if
$\epsilon_Q$ vanishes and also if the K\"ahler potential respects
$Z_2^{X\leftrightarrow X'}$ symmetry. In this case, we avoid the
constraint from the EDMs, and $\epsilon_1$ can be much larger than
$\epsilon_A$. For example, if the $Z_2^{X\leftrightarrow X'}$ symmetry
breaking field $\phi$ and the Yukawa coupling $y_1$ have a non-trivial
transformation property under some symmetry (like $R$-symmetry),
$\epsilon_A$ is expected to be $O(y_1^2\epsilon_1)$, which can be
suppressed for smaller $y_1$.

In our discussion, we assumed that there is no effect of the
$Z_2^{X\leftrightarrow X'}$ symmetry breaking in the gauge kinetic
function. If there is such an effect for the strong gauge groups
SU(2)$_{\rm S}$ and ${\rm SU(2)'_{S}}$, the relative phase of
$\Lambda$ and $\Lambda'$ becomes $O(8\pi^2\epsilon/g_{\rm S}^2)$,
where $g_{\rm S}$ is the gauge coupling constant for the strong gauge
groups.  Therefore, small symmetry breaking effect may induce a large
shift of the relative phase of $F_X$ and $F_{X'}$, resulting in a
large $\epsilon_A$.\footnote
 {This may not happen if the coupling constants for the strong gauge
groups become non-perturbative at the Planck scale.}
 Therefore, the $Z_2^{X\leftrightarrow X'}$ symmetry breaking in the
gauge kinetic function is disfavored. This effect can also be killed
if non-trivial transformation properties for some symmetry are
assigned for the symmetry breaking parameters.\footnote
 {Contrary to the strong gauge groups, there may be an effect of the
$Z_2^{X\leftrightarrow X'}$ symmetry breaking in the balancing gauge
groups sector (${\rm SU(2)_B}$'s), since our result is not affected by
the phases of the strong scales of these interactions. Of course, the
hierarchy between $v$ and $v'$ is affected by this effect.}

\section{Summary}

In the first half of this letter, we calculated the EDM of the
electron in the framework of the gauge mediated model. If all the
phases in the Lagrangian are $O(1)$, the electron EDM is larger than
the current experimental constraint. If all the superpaticles have
masses of $O({\rm 100~GeV})$, for example, the CP violating phase
$\theta_{\rm phys}$ has to be smaller than $O(10^{-2})-O(10^{-3})$
depending on $\tan\beta$.

Regarding this tuning as a problem, we considered a mechanism to
suppress the CP violating phase. If the $\mu$- and $B_\mu$-parameters
originate from the same coupling to the SUSY breaking field in the
superpotential, the physical phase is cancelled out. However, the
ratio $B_\mu/\mu$ becomes too large in a naive model. Therefore, we
introduced another sector to suppress this ratio.  Even with the new
field, we have seen that the smallness of the physical phase
$\theta_{\rm phys}$ can be realized by a symmetry.

Finally, we note that the strong CP problem cannot be solved in our
model. This feature is common to the case of the SSM, and some
mechanism is needed to solve this problem, like Peccei-Quinn
symmetry~\cite{PecQui}.

\section*{Acknowledgment}

The author would like to thank J.~Bagger for stimulating discussions.
He is also grateful to J.L.~Feng for useful comments and careful
reading of the manuscript. This work was supported by the National
Science Foundation under grant PHY-9513835.

\end{document}